\documentclass[aps,prc,preprintnumbers,showpacs,showkeys,nofootinbib,
superscriptaddress,fleqn,floatfix,tightenlines,10pt,
twocolumn,byrevtex]{revtex4-1} 
\usepackage{amsmath,amsfonts,amssymb,amscd,amsxtra,amsthm}
\usepackage{graphicx}  
\usepackage{epstopdf}
\usepackage{dcolumn}  
\usepackage{bm}          
\usepackage{slashed}
\usepackage{cancel} 
\usepackage{float} 
\usepackage{mathtools}
\usepackage{amsbsy}
\usepackage{amstext}

\usepackage[utf8]{inputenc} 
\usepackage[normalem]{ulem} 
\usepackage[dvipsnames]{xcolor} 
\renewcommand\sout{\bgroup \color{red} \ULdepth=-.5ex \ULset}

\newcommand\T{\rule{0pt}{2.6ex}}       

\makeatletter

\begin{document}
\preprint{INHA-NTG-06/2018}
\title{Modification of hyperon masses in nuclear matter}
\author{Ki-Hoon Hong}
\email[E-mail: ]{kihoon@inha.edu}
\affiliation{Department of Physics, Inha University, Incheon 22212,
 Republic of  Korea}
\author{Ulugbek Yakhshiev}
\email[E-mail: ]{yakhshiev@inha.ac.kr}
\affiliation{Department of Physics, Inha University, Incheon 22212,
 Republic of Korea}
\author{Hyun-Chul Kim}
\email[E-mail: ]{hchkim@inha.ac.kr}
\affiliation{Department of Physics, Inha University, Incheon 22212,
  Republic of Korea}
\affiliation{Advanced Science Research Center, Japan Atomic Energy
  Agency, Shirakata, Tokai, Ibaraki, 319-1195, Japan}
\affiliation{School of Physics, Korea Institute for Advanced Study 
 (KIAS), Seoul 02455, Republic of Korea}  
\begin{abstract}
We investigate the properties of baryons within the framework of the
in-medium modified SU(3) Skyrme model.  The modification is performed
by a minimal way, the medium functionals in the SU(2) sector being
introduced. These functionals are then related to nuclear matter
properties near the saturation point. The modifications in the SU(3)
sector are performed by changing additionally kaon properties in
nuclear matter. The results show that the properties of baryons in
the strange sector are sensitive to the in-medium modifications of
the kaon properties. We discuss the consistency of the in-medium
modifications of hadron properties in this approach, comparing
the present results with those from other models.
\end{abstract}


\date {\today}
\maketitle

\section{Introduction}
 Nucleons are known to undergo changes in nuclear matter
due to the strong interaction with nuclear environment. Since 
they themselves constitute nuclear matter, the medium modifications of
nucleon properties bring about the changes of nuclear matter in a 
self-consistent manner. Similarly, a hyperon lying in nuclear matter
is also altered. It is essential to understand how its attributes
become different in nuclear medium so that neutron stars and 
hypernuclei can be described in a more realistic
way~\cite{Balberg:1998ug,Djapo:2008au, Demorest:2010bx,
  Bednarek:2011gd,Katayama:2015dga, Lattimer:2015nhk, Gal:2016boi,
  Oertel:2016bki}.   

While a plethora of experimental and theoretical works on conventional
nonstrange nuclear matter and its constituents in a wide range of
nuclear matter densities has been compiled well over decades, hyperons
in nuclear matter have been relatively less
studied~\cite{Saito:1994kg,Schaffner:1995th, Savage:1995kv,  
  Beane:2012ey, Haidenbauer:2014uua,Petschauer:2015nea,
  Lenske:2018bgr}. Most of the works are based on the
hyperon-nucleon ($YN$) interactions. For example, Beane et
al.~\cite{Beane:2012ey} computed the $n\Sigma^-$ scattering phase
shifts using lattice QCD to quantify the energy shift of the
$\Sigma^-$ in nuclear
matter. In Refs.~\cite{Haidenbauer:2014uua,Petschauer:2015nea}   
the $YN$ potential was constructed from effective field theory and
the Bruecker-Hartree-Fock (BHF) approximation was employed to
investigate hyperons in nuclear matter. Density functional theories
were also used to study the hyperons in nuclear matter (see a recent
review~\cite{Lenske:2018bgr}).     

In the present work, we propose yet another simple framework of
investigating the mass shifts of the hyperons together with the
nucleon and the $\Delta$ isobar. Some years ago, it was already studied
how they underwent the changes in nuclear matter within the framework
of the chiral topological soliton models~\cite{Kim:2012ts,
  Jung:2012sy, Yakhshiev:2013goa, Jung:2014jja, Jung:2015piw}, where
the mass shifts of the nonstrange baryons were scrutinized and various
in-medium modified form factors were computed. The 
results were in qualitative agreement with those of other
approaches and had interesting physical implications such as the 
stability and shape of the nucleon in nuclear medium. 

Moreover, the in-medium modified SU(2) Skyrme model described
very well properties of isospin asymmetric nuclear matter near the
saturation point (nuclear density $\rho_0=0.16$\,fm$^{-3}$). 
The model yielded successfully the equations of states (EoS) for
nuclear matter at ordinary densities~\cite{Yakhshiev:2013eya}.  
It predicted qualitatively various properties of nuclear matter in
comparison with different theoretical approaches and empirical
information. In particular, the parameters of the symmetric and
asymmetric EoS determined from the present model were in qualitative
and quantitative agreement with the empirical data~\cite{Li:2007bp},
with those from Hartree-Fock approaches based  on Skyrme
 interactions~\cite{Chabanat:1997qh,Trippa:2008gr} and with those 
from different aproaches presented in Refs.~\cite{Chen:2009qc,
  Sharma:1988zza,Shlomo:1993zz,Li:2005jy}. Furthermore, the 
extrapolations of the EoS at higher densities indicate that the  
model can describe rather well the state of matter that may exist in
the interior of neutron stars. The results demonstrate that two solar
mass neutron stars can be explained in the framework 
of the present approach~\cite{Yakhshiev:2015noa}.    

In this context, it is of great interest and significance to extend
the SU(2) version of the model to the SU(3) one in a straightforward
and simple manner. So, we will generalize the previous analyses
to investigate the hyperons in nuclear matter. We will employ an SU(3)
Skyrme model developed in Ref.~\cite{Westerberg:1994ef} and modify the
relevant parameters of the model in nuclear matter. For simplicity, we
first consider only the in-medium modification of meson dynamics in
the SU(2) sector.  However, the kaon is also known to undergo the
changes in nuclear matter~\cite{Waas:1996xh, Waas:1996tw}. Thus, we 
alter the kaon properties in nuclear medium, assuming a simple
linear-density approximation. While the dynamics in the SU(2) sector
remained intact in the course of generalization to the SU(3) sector,
the model still properly explains the phenomenology in the nonstrange
sector as discussed in
Refs.~\cite{Yakhshiev:2013eya,Yakhshiev:2015noa}.  
The present approach allows one to draw a simple conclusion as to how  
the in-medium modified kaon can influence the changes of the SU(3)
baryons in nuclear matter. 

The paper is organized as follows: In Section II, we recapitulate
briefly an SU(3) Skyrme model in free space~\cite{Westerberg:1994ef}, 
where the SU(2) Skyrme model is extended into the SU(3) by a trivial
embedding of the SU(2) chiral soliton into SU(3)~\cite{Witten:1983tx}.  
In addition, we show how the strange sector incorporates the quantum 
fluctuations (see subsection~\ref{subsec:BarInFS}).  
Then we explain how the meson dynamics is altered in nuclear medium,  
based on the phenomenology in the nonstrange sector and then 
discuss the in-medium changes of nucleon and $\Delta$
isobar properties in the subsection~\ref{subsec:BarInNM}. The
modification of kaon properties in nuclear matter is discussed in the
subsection~\ref{subsec:KaonInNM}. In Section~\ref{sec:ResDis} we
present and discuss the results. We first deal with the medium 
effects in the mesonic sector (subsection~\ref{subsec:inm-messec}) 
and then we show how the modification of the mesonic sector brings
about the density effects on the hyperons
(subsection~\ref{subsec:inm-barsec}). The final
Section~\ref{sec:Sum&Out} is devoted to the summary of the present
work and outlook of possible developments of the present model in
relation with the strangeness physics in various nuclear environments.   
 
\section{The model}
The SU(3) Skyrme models have been developed over decades. There are 
many variants of the model~\cite{Guadagnini:1983uv, Mazur:1984yf,
  Chemtob:1985ar, Callan:1985hy, Callan:1987xt, Yabu:1987hm,
  Weigel:1995cz} (see a review~\cite{Weigel:2008zz} for extensive
references). The main difference among the models comes mainly
from specific methods as to how the strange sector is treated. We will
follow an SU(3) Skyrme model developed in
Ref.~\cite{Westerberg:1994ef}, because one can easily  
and transparently modify the model in nuclear medium.   

\subsection{Baryons in free space}
\label{subsec:BarInFS}
The standard SU(3) Skyrme model is based on the effective chiral 
Lagrangian written by 
\begin{align}
  \mathcal{L}&=\mathcal{L}_{\mathrm{WZ}}-\frac{F_\pi^2}{16}{\rm Tr}\,
               L_\mu L^\mu  +\frac{1}{32e^2}{\rm
               Tr}\,[L_\mu,L_\nu]^2\cr 
 & +\frac{F_\pi^2}{16}{\rm Tr}\,\mathcal{M}(U+U^\dagger-2),
\end{align}
where  $L_\mu=U^\dagger\partial_\mu U$ and $U(\bm{x},t)$ is a chiral field
in SU(3). The mass matrix ${\mathcal M}$ is defined in terms of the pion
and kaon masses  
\begin{align}
\mathcal{M}=\left(
\begin{array}{ccc}
m_\pi^2 &0&0\\
0&m_\pi^2 &0\\
0&0&2m_K^2-m_\pi^2
\end{array}
\right),
\end{align}
where $m_\pi$ and $m_K$ stand for the pion and kaon masses,
respectively. 
The Wess-Zumino term~\cite{Wess:1971yu}
$\mathcal{L}_{\mathrm{WZ}}$ constrains the soliton to identify as a
baryon, which is expressed by the five-dimensional integral over a
disk $D$  
\begin{align}
S_{\rm WZ} = -\frac{iN_c}{240\pi^2} \int_{D} d^5 \vec x\,
   \epsilon^{\mu\nu\alpha\beta\gamma}
   {\rm Tr}(L_\mu L_\nu L_\alpha L_\beta L_\gamma).
\end{align}
Here the totally antisymmetric tensor
$\epsilon^{\mu\nu\alpha\beta\gamma}$ is defined as 
$\epsilon^{01234}=1$ and $N_c=3$ is the number of colors. The input
parameters of the model are the pion decay constant 
$F_\pi=108.783$\,MeV, the Skyrme parameter $e=4.854$, and the masses
of the $\pi$ and $K$ mesons, given respectively as
$m_\pi=134.976$\,MeV and $m_K=495$\,MeV, which are taken close 
to the experimental data.   

Classically, the model describes a set of absolutely stable
topological solitons with the corresponding topological integer 
numbers that is identified as a baryon number $B$. 
The lowest-lying baryon states can be obtained by the zero-mode
quantization of the soliton with baryon number $B=1$
\begin{equation}
U(\bm{r},t)=\mathcal{A}(t)U_0(\bm{r})\mathcal{A}(t)^\dagger,
\end{equation}
where $\mathcal{A}(t)$ is rotational matrix in SU(3). The
time-independent soliton field $U_0(\bm{r})$ is expressed as the
trivial embedding of the SU(2) soliton into SU(3)
\begin{align}
U_0(\bm{r})=\left(
\begin{array}{cc}
e^{i\bm{\tau}\cdot\bm{n}F(r)}&0\\
0&1\end{array}\right),\qquad \bm{n}=\frac{\bm{r}}{r}\,.
\label{StatAnz}
\end{align}
Note that the SU(2) soliton field satisfies the hedgehog ansatz.
The profile function $F(r)$ with the boundary conditions 
\begin{align}
F(0)=\pi,\qquad F(\infty)=0
\label{BoundCon}
\end{align}
satisfies the classical field equations corresponding to the baryon
number $B=1$ solution. 

The model~\cite{Westerberg:1994ef} is characterized in dealing with
the time-dependent rotational matrix $\mathcal{A}(t)$. While the SU(2)   
rotation is restricted to the nonstrange sector represented by $A(t)$,
the transformation along the strange sector is governed by the new
matrix $S(t)$. Thus, $\mathcal{A}(t)$ and $S(t)$ can be
expressed respectively as   
\begin{align}
\mathcal{A}(t)&=\left(\begin{array}{cc}
A(t)&0\\
0^\dagger&1\end{array}\right)S(t),\\
A(t)&=k_0(t){\bf 1}+i \sum_{a=1}^3\tau_a k_a(t),\\
  S(t)&=\exp\left\{i\sum_{p=4}^7k_p \lambda_p\right\}\cr
  &\equiv\exp\left(i\mathcal{D}\right)
  =\exp\left\{\left(\begin{array}{cc}
    0&i\sqrt{2}D\\
  i\sqrt{2}D^\dagger&0\end{array}\right)
  \right\},
  \label{Smatrix}
\end{align}
with 
\begin{align}
  D(t)&=\frac{1}{\sqrt{2}}\left(\begin{array}{c}
  k_4(t)-ik_5(t)\\
k_6(t)-ik_7(t)\end{array}\right).
\end{align}
Here $\tau_{1,2,3}$ denote the Pauli matrices, whereas $\lambda_p$
stand for the strange part of the SU(3) Gell-Mann matrices.  
$k_a(t)$ $(a=0,1,2,\dots,7)$ represent arbitrary collective
coordinates.  The matrix $A(t)$ with the collective
coordinates $k_a$  $(a=0,1,2,3)$ stand for the rotational fluctuation
of the SU(2) static soliton in the nonstrange sector. On the other
hand, the matrix $S(t)$ with the collective coordinates $k_p$
$(p=4,5,6,7)$ describes the zero-mode fluctuation along the strangeness 
direction. Note that the Wess-Zumino term imposes a constraint on the
eighth component which is related to the baryon number. 

$S(t)$ in Eq.\,(\ref{Smatrix}) can be systematically expanded in terms of 
matrix $D(t)$ because $\mathcal{D}(t)$ satisfies the relation
\begin{align*}
\mathcal{D}^3=d^2\mathcal{D},\qquad d^2\equiv2D^\dagger D.
\end{align*}
We will perform the expansion and will keep lower orders in power
of $D$ terms (including $D^4$ terms) in the Lagrangian. 
Having expanded $S(t)$, we obtain the time-dependent Lagrangian in the
form 
\begin{align}
  L=&-E_0+4\Phi \dot{D}^\dagger\dot{D}-\Gamma 
  M^2D^\dagger D\cr
  &+\frac{iN_c}{2}(D^\dagger\dot{D}-\dot{D}^\dagger D) +
    \frac{1}{2}\Omega\omega^2\cr 
  &+i(\Omega-2\Phi )(D^\dagger(\bm{\omega}\cdot\bm{\tau})\dot{D}
  -\dot{D}^\dagger(\bm{\omega}\cdot\bm{\tau})D)\cr
&  +2\left(\Omega-\frac{4}{3}\Phi \right)(D^\dagger D)
  (\dot{D}^\dagger\dot{D})\cr 
  &-\frac{1}{2}\left(\Omega-\frac{4}{3}\Phi \right)
    (D^\dagger\dot{D}+\dot{D}^\dagger D)^2\cr 
  &+2\Phi (D^\dagger\dot{D}-\dot{D}^\dagger D)^2
  +\frac{2}{3}\Gamma M^2(D^\dagger D)^2\cr
  &-\frac{N_c}{2}(D^\dagger(\bm{\omega}\cdot{\bm{\tau}})D)
  \cr  &
  -\frac{iN_c}{3}(D^\dagger D)(D^\dagger\dot{D}-\dot{D}^\dagger D), 
\end{align}
where  $M^2=m_K^2-m_\pi^2$ and $\bm{\omega}$ denotes the rotational
velocity in SU(2), defined by 
\begin{align}
A^\dagger\dot A=\frac{1}{2}\,\bm{\omega}\cdot\bm{\tau}.
\end{align}
The energy of the static configuration $E_0[F]$ is derived as 
\begin{align}
  E_0[F]&=4\pi\int_0^\infty dr\, r^2\Big\{\frac{F_\pi^2}{8}
          \left(\frac{2\sin^2F}{r^2} 
  +F_r^2\right)\cr
  &+\frac{1}{2e^2}\frac{\sin^2F}{r^2}
  \left(\frac{\sin^2F}{r^2}+2F_r^2\right)\cr
  &+\frac{F_\pi^2m_\pi^2}{2}\sin^2\frac{F}{2}\Big\},
\end{align}
where $F_r\equiv\partial_rF$.
Minimizing this functional, we obtain the solutions of the field
equations with the boundary conditions defined in
Eq.~(\ref{BoundCon}). 

The functional $\Omega[F]$ arises from the rotations of the static
soliton in the SU(2) sector, whereas the functionals $\Phi[F]$ and
$\Gamma[F]$ explain the deviation into the strangeness sector. They 
are expressed as 
\begin{align}
  \Omega[F]&=\frac{2\pi}{3}\int_0^\infty dr\, r^2\sin^2F\cr
  &\times\Big\{F_\pi^2+\frac{4}{e^2}
    \left(F_r^2+\frac{\sin^2F}{r^2}\right)\Big\},\\ 
  \Phi[F]&=\pi\int_0^\infty dr\, r^2\sin^2\frac{F}{2}\cr
  &\times\Big\{F_\pi^2+\frac{1}{e^2}\left(F_r^2
    +\frac{2\sin^2F}{r^2}\right)\!\!\Big\},\\ 
  \Gamma[F] &=4\pi\int_0^\infty dr\, r^2 F_\pi^2\sin^2\frac{F}{2}. 
\end{align}
 
In order to quantize the soliton, one introduces the canonical
momenta conjugate to the $\omega_i$ and $\dot{D}$, which   
correspond respectively to the SU(2) rotation and the deviation to the
strangeness direction 
\begin{align}
  (J_{ud})_i&=\frac{\partial L}{\partial\omega^i}\,,\qquad
  \Pi^\gamma=\frac{\partial L}{\partial\dot{D}^\dagger_\gamma}\,.
\end{align}
They satisfy the following commutation relations
\begin{align}
  [(J_{ud})_i,\alpha^j]&=\frac{1}{i}\delta^j_i\\
  [\Pi^\gamma,D^\dagger_\beta]&=[\Pi^\dagger_\beta,D^\gamma] =
                                \frac{1}{i}\delta^\gamma_\beta. 
\end{align}
The angular momentum operator $\bm{J}_{ud}$ and the momentum $\Pi$ are
derived as 
\begin{align}
  \bm{J}_{ud}&=\Omega\bm{\omega}+i(\Omega-2\Phi)
               (D^\dagger\bm{\tau}\dot{D}-\dot{D}^\dagger\bm{\tau}D)\cr  
 & -\frac{N_c}{2}D^\dagger\bm{\tau}D,
 \label{Jud}\\
  \Pi&=4\Phi\dot{D}-\frac{iN_c}{2}D-i(\Omega-2\Phi)\bm{\omega}
       \cdot\bm{\tau}D\cr
       &-\left(\Omega-\frac{4}{3}\right)(D^\dagger\dot{D}+\dot{D}^\dagger D)D\cr
 &-4\Phi(D^\dagger\dot{D}-\dot{D}^\dagger D)D+\frac{1}{3}N_c
 (D^\dagger D)D\cr
 &+2\left(\Omega-\frac{4}{3}
 \Phi\right)(D^\dagger D) \dot{D}.
\end{align}

Then we obtain the collective Hamiltonian to order $N_c^{-1}$ as
follows 
\begin{align}
  H
  &=E_0+\frac{1}{4\Phi}\Pi^\dagger\Pi+\left(\Gamma  M^2
  +\frac{N^2}{16\Phi}\right)D^\dagger D\cr
  &-\frac{iN_c}{8\Phi}(D^\dagger\Pi-\Pi^\dagger D)
  +\frac{1}{2\Omega}\bm{J}_{ud}^2+\frac{N_c}{4\Phi}D^\dagger 
  \bm{J}_{ud}\cdot\bm{\tau}D\cr
  &+i\left(\frac{1}{2\Omega}-\frac{1}{4\Phi}\right)
  (D^\dagger\bm{J}_{ud}\cdot\bm{\tau}\Pi-
  \Pi^\dagger\bm{J}_{ud}\cdot  \bm{\tau}D)\cr 
&  \left(\frac{1}{2\Omega}-\frac{1}{3\Phi}\right)
\left((D^\dagger D)(\Pi^\dagger\Pi)-\frac{1}{4}
(D^\dagger\Pi+\Pi^\dagger D)^2\right)\cr
  &-\frac{1}{4\Phi}(D^\dagger\Pi-\Pi^\dagger D)^2-i\frac{N_c}{8\Phi}
  (D^\dagger\Pi-\Pi^\dagger D)D^\dagger D\cr
  &+\left(\frac{N_c^2}{12\Phi}-\frac{2}{3}\Gamma m_K^2\right)
  (D^\dagger D)^2.
  \label{Hgeneral}
\end{align}
The collective Hamiltonian can be diagonalized by introducing
the creation and annihilation operators instead of $D$ and $\Pi$
\begin{align}
  D=&\frac{1}{\sqrt{N_c}}\left(1+\frac{ M^2}{M_0^2}\right)^{-1/4}
      (a+b^\dagger)\\ 
  \Pi=&-\frac{i}{2}\sqrt{N_c}\left(1+\frac{ M^2}{M_0^2}
        \right)^{1/4}(a-b^\dagger), 
\end{align}
where $M_0$ is defined as $M_0={N_c}/(4\sqrt{\Phi \Gamma})$.  The
operators  $a^\dagger (a)$ and $b^\dagger(b)$ denote respectively the
creation (annihilation) operators of the strange quark and antiquark,
respectively. The strangeness and the angular momentum of the strange 
quark are given respectively by 
\begin{align}
  s=b^\dagger b-a^\dagger a, \qquad\bm{J}_s =
  \frac{1}{2}(a^\dagger\bm{\tau}a - b\bm{\tau}b^\dagger).
\end{align}
Then the normal-ordered Hamiltonian to order $N_c^0$ is derived as 
\begin{align}
  H=E_0+\omega_{-} a^\dagger a+\omega_{+}b^\dagger b,
\end{align}
where
\begin{align}
\omega_{\pm}=\frac{N_c}{8\Phi
  }\left(\sqrt{1+\frac{16\Phi\Gamma}{N_c^2}\,M^2}\pm 1\right). 
\end{align}
Since we are interested in baryons containing only the strange quarks, 
not antiquarks, we will ignore the $\omega_+$ term in the
Hamiltonian. Furthermore, we also neglect the quartic terms in the
kaon field because classical dynamics in the mesonic sector is still 
restricted to the pion-pion interaction. 
Thus, ignoring the corresponding terms related to the kaon-kaon
interaction in Eq.~\eqref{Hgeneral}, we arrive at the final
expressions of the collective Hamiltonian   
\begin{align}
  H&=E_0+\omega_{-} a^\dagger a+\frac{1}{2\Omega}
     \left(\bm{J}_{ud}+c\bm{J}_s\right)^2, 
\end{align}
where $c$ is defined as
\begin{align}
  c=1-  \frac{4\Omega\omega_{-}}{8\Phi\omega_{-}+N_c}. 
\end{align}

Sandwiching the collective Hamiltonian between the eigenstates $
|n_s\rangle|I,J\rangle$ with the definite quantum numbers such as isospin
$I$, total angular momentum $J$ and given number of strange quarks,     
we obtain the final mass formula of the SU(3) baryons
\begin{align}
M&=E_0-s\omega_{-}+\frac{1}{2\Omega}\Big\{cJ(J+1)\cr
&+(1-c)I(I+1)+\frac{c(c-1)}{4}\,s(s-2)\Big\}.
\label{eq:mass_free}
\end{align}
More details of the model in free space can be found in
Refs.~\cite{Westerberg:1994ef,Kaplan:1989fc}.  

\subsection{Baryons in nuclear matter}
\label{subsec:BarInNM}
We now show how to implement the medium effects into the SU(3) Skyrme
model. For simplicity, we will first take into account a modification
of meson dynamics in the SU(2) sector, introducing the medium  
functionals into the effective chiral Lagrangian, based on the
low-energy phenomenology in nuclear medium~\cite{Ericsonbook}. 
As we mentioned already, the SU(2) Skyrme model was parametrized in
terms of the density functionals and was applied successfully to the
description of properties of the nucleon and $\Delta$ isobar near the
normal nuclear matter density $\rho_0$~\cite{Yakhshiev:2013eya}. The
model was even well extrapolated to higher density
regions~\cite{Yakhshiev:2015noa}.  

In Ref.~\cite{Yakhshiev:2013eya}, the in-medium modified SU(2) Skyrme 
model was discussed in detail, isospin symmetric and asymmetric
infinite nuclear matter being considered. The effective chiral
Lagrangian is modified as follows 
\begin{align}
  \mathcal{L}=&-\frac{F_\pi^2}{16}\alpha_2^t(\rho)
  {\rm Tr} L_0L_0+\frac{F_\pi^2}{16}\alpha_2^s(\rho){\rm Tr} L_iL_i\cr
  &-\frac{\alpha_4^t(\rho)}{16e^2}
  {\rm Tr}[L_0,L_i]^2 +\frac{\alpha_4^s(\rho)}{32e^2}{\rm Tr}[L_i,L_j]^2\cr
&  +\frac{F_\pi^2}{16}\alpha_{\chi SB}(\rho){\rm Tr} \mathcal{M}(U+U^\dagger-2),
\label{ModLag}
\end{align} 
where $\alpha_2^t(\rho)$, $\alpha_2^s(\rho)$, $\alpha_4^t(\rho)$,
$\alpha_4^s(\rho)$ and $\alpha_{\chi SB}(\rho)$ denote the functionals
of the nuclear matter density, which reflect the changes of meson
properties in nuclear medium. 
In principle, they should be defined in a self-consistent
way. However, it will be extremely difficult to determine them
self-consistently, in particular, when one considers real nuclei with
respect to their in-medium modified constituents. Therefore, we simply
assume these medium functionals to be external functions of nuclear
matter density $\rho$. Then we are able to study properties of a
single baryon in nuclear matter. This 
assumption is a rather plausible one, as far as we are interested in
homogenous infinite nuclear matter. The medium functionals can indeed
be considered as simple external parameters at a given density so
that one can carry out the calculations in a easy manner. Furthermore,
the density-dependent parameters can be related to the properties of 
infinite nuclear matter, so that one can partially restore the
self-consistency of the model~\cite{Yakhshiev:2013eya}.  

In the present work, we will generalize the method developed in
Ref.~\cite{Yakhshiev:2013eya}. The in-medium modified Lagrangian in
SU(3) will be modified as done in Eq.\,(\ref{ModLag}), the Wess-Zumino
Lagrangian ${\cal L}_{WZ}$ being included. However, we note that
the Wess-Zumino term should not be modified in nuclear matter, since
the topology of the model must be kept intact such that the
baryon number is preserved. So, the Wess-Zumino term is modified in
nuclear matter only inexplicitly through the medium modification of
the solutions with the same baryon number in nuclear matter. 

We want to mention an important aspect of the present approach.  
The Skyrme model is based on a truncated version of the most general
effective chiral Lagrangian. It indicates that the contributions from 
higher-order terms enter tacitly into the parameter of the Skyrme
term. This means that the parameter carries the effects of the
higher-order contributions effectively. Therefore, the in-medium
modified Skyrme model keeps already almost all the necessary   
ingredients and in principle could be a relevant theoretical framework  
to study nuclear many-body problems, at least to a qualitative
extent.  For example, the in-medium Skyrme term plays an essential role in
stabilizing the nucleon even in nuclear matter. The Skyrme term brings
about the repulsive nature in the inner part of the
nucleon~\cite{Kim:2012ts, Jung:2014jja}, which assures the stability
of the nucleon. It implies that when the density of nuclear matter
grows higher-order terms of the effective chiral Lagrangian will
definitely come into play and are required so that the collapse of
nuclear matter to a singularity~\cite{Yakhshiev:2010kf} be avoided. 
Therefore, the effect of higher order derivative terms is incorporated 
by introducing the density-dependent parameter in the Skyrme term.

If the functionals are taken to be functions of nuclear-matter density 
$\alpha(\rho)$, then all the functional parameters are reduced to the 
simple external parameters in an infinite and homogeneous nuclear
matter approximation. We will follow in this work the method developed
in a previous work~\cite{Yakhshiev:2013eya}. First, we introduce a
\emph{convenient} relation between the medium functions in the
following way   
\begin{equation}
 \alpha_2^t=\alpha_4^t\alpha_2^s(\alpha_4^s)^{-1},
 \label{relat}
\end{equation}
which reduces the number of the external density-dependent parameters
to four different parameters. Furthermore, these remaining four
density-dependent parameters can be related to each other, so that the
three independent parameters can be defined as follows 
 \begin{align}
1+C_1\lambda=&f_1(\lambda)\equiv\sqrt{\alpha_2^s\alpha_4^s},
\label{f1}\\
1+C_2\lambda=&f_2(\lambda)\equiv\sqrt{\frac{\alpha_{\chi SB}
               \alpha_4^s}{(\alpha_2^s)^2}},\\ 
1+C_3\lambda=&f_3(\lambda)\equiv\left(\frac{\alpha_2^s}{\alpha_4^s}
               \right)^{\frac{3}{2}}  \frac{1}{\alpha_2^t},
 \label{f3}
 \end{align} 
where $\lambda=\rho/\rho_0$. This reduction allows to keep the medium
modification of the parameters simpler and more general. 

By defining Eq.~(\ref{relat}) and by introducing
Eqs.~(\ref{f1})-(\ref{f3}), an algebraic manipulations become much
simplified and yields convenient and transparent forms of the final
expressions. Then we can perform the main part of calculations such as
the minimization, the quantization, and so on, in terms of the three
independent density-dependent functions $f_{1,2,3}$.  

The numerical values of these parameters are fixed to be
$C_1=-0.279$, $C_2=0.737 $ and $C_3=1.782$.  They reproduce well the 
EoS for symmetric nuclear  matter near $\rho_0$ and at higher
densities that may exist in the interior of a neutron
star~\cite{Yakhshiev:2013eya,Yakhshiev:2015noa}.   
The parameters of the present model are completely fixed in nuclear
matter except for the strangeness direction. So, we will then
introduce the modification of the kaon properties after the
quantization, which will be discussed in
subsection~\ref{subsec:KaonInNM}. 

While the form of the baryon mass formula is kept to be the same as in
Eq.~\eqref{eq:mass_free}, it becomes now density-dependent by the
functions $f_{1,2,3}$ \footnote{Quantities with the asterisk  ``*''
in the expressions stand for those modified in nuclear medium in terms of the
density-dependent functions $f_{1,2,3}$.}   
\begin{align}
M^*&=E_0^*-s\omega^*_{-}+\frac{1}{2\Omega^*}\left[c^*J(J+1) \right.\cr 
&\left. \!+(1-c^*)I(I+1)+\frac{c^*(c^*-1)}{4}s(s-2)\! \right],
\label{massformula}
\end{align}
where $\omega^*_{-}$ and $c^*$ are changed as 
\begin{align}
  \omega^*_{-}&=\frac{N_c}{8\Phi^*}\left(\sqrt{1+\frac{16\Phi^*\Gamma^*}
  {N_c^2}\,M^{*2}}-1\right),\\
  c^*&=1-\frac{4\Omega^*\omega^*_{-}}{8\Phi^*\omega^*_{-}+N_c},
  \label{omegastar}
\end{align}
where $M^{*2}=m_K^{*2}-m_\pi^2$.
In Eq.~\eqref{omegastar} the value of the kaon mass is released from
the experimental data in free space by considering the medium
effects. The classical soliton mass $E_0^*$, $\omega^*_{-}$, and $c^*$
are expressed respectively as  
\begin{align}
 E_0^*&=f_1\frac{4\pi F_{\pi}}{e}\int_0^\infty dx\, x^2\Big\{
 \frac{1}{8}\left(F_x^2+\frac{2\sin^2F}{x^2}\right)\cr
 &+\frac{\sin^2F}{x^2}\left(F_x^2+\frac{\sin^2F}{2x^2}\right)
 +\frac{\beta^2}{2}\sin^2\frac{F}{2}\Big\},\\
 \Omega^*&= f_3^{-1}\frac{2\pi}{3e^3F_\pi}\int_0^\infty dx\, x^2\sin^2F\cr
 &\times\Big\{1 +4\left(F_x^2+\frac{\sin^2F}{x^2}\right)\Big\},\\
 \Phi^*&=f_3^{-1}\frac{\pi}{e^3F_\pi }\int_0^\infty dx\, x^2\sin^2\frac{F}{2}\cr
 &\times \Big\{1
+\left(F_x^2+\frac{2\sin^2F}{x^2}\right)\Big\},\\
 \Gamma^*&=f_1f_2^2\frac{4\pi}{e^3F_\pi}
 \int_0^\infty dx\, x^2\sin^2\frac{F}{2},
\end{align}
where we have introduced a parameter $\beta=f_2{m_\pi}/{eF_\pi}$  and
a dimensionless variable 
$x=eF_\pi({\alpha_2^s}/{\alpha_4^s})^{1/2}r$. 
Other aspects of the medium modifications can be found in
Ref.~\cite{Yakhshiev:2013eya} and references therein.

\subsection{Kaon properties in nuclear matter}
\label{subsec:KaonInNM}
We are now in a position to deal with the change of kaon properties in
nuclear matter. As seen in Eq.~\eqref{omegastar} an additional medium
modification was implemented by the kaon mass in nuclear matter. 
Before we carry on the explicit calculation of the SU(3) baryon masses 
in nuclear matter, we need to explain how the kaon properties undergo
the change in nuclear environment. To be more consistent, one should
consider how the kaon propagator is altered in nuclear matter, which
arises from the polarization effects, as done for that of the pion in 
nuclear matter~\cite{Ericsonbook}. The polarization operator can be   
described phenomenologically by introducing a kaon-nucleus optical
potential. The properties of this optical potential may be related
either to the phenomenology of kaon-nucleus scattering or to the
properties of kaonic atoms as done for the non-strange sector (see
e.g. Ref.\,\cite{Yakhshiev:2013eya} and references therein). Because
of a lack of the experimental data, it is however rather
difficult to extract information on how the kaon properties are
varied in nuclear matter. Thus, instead of conducting such a
complicated analysis, we will rather take into account a simple
modification of the kaon properties after the quantization in the
present work, keeping dynamics of the mesonic sector intact in nuclear
medium. Since it is known that the kaon mass drops off in dense  
matter~\cite{Kaplan:1987sc, Kolomeitsev:1995xz}, we will consider only
the change of the kaon mass in the present work as a minimal
modification of the kaon properties in nuclear matter. 

Here we note, that the in-medium modified Lagrangian in 
Eq.~\eqref{ModLag} can be
reformulated in terms of the in-medium modified pion decay constants
and the Skyrme parameters
\begin{align}
F_{\pi,t}^*&=F_\pi\sqrt{\alpha_2^t}, \quad
F_{\pi,s}^*=F_\pi\sqrt{\alpha_2^s},\cr
 e_t^*&=\frac{e}{\sqrt{\alpha_4^t}},\quad
e_s^*=\frac{e}{\sqrt{\alpha_4^s}},
\label{medpar}\\
m_\pi^*&=m_\pi\sqrt{\frac{\alpha_{\chi SB}}{\alpha_2^s}}\,.\nonumber
\end{align}
Then the change of SU(2) dynamics in nuclear matter can be understood as
the medium modification of the input parameters. 

In the SU(3) case, we need to modify the kaon decay constant and the
kaon mass in addition to the pion observables. Within the present
framework, the kaon decay constant $F_K$ is assumed to be equal to the
pion decay  constant $F_\pi$ in free space. This is a reasonable
consideration, though the value of $F_K$ is larger than that of
$F_\pi$. At least, these two constants become equal in the SU(3)
symmetric case. If we assume that the modified kaon decay constant
will have exactly the same form of the pion decay constant in nuclear
medium $F_K^*=F_{\pi,s}^*$, Eq.~\eqref{medpar} implies that the kaon
mass may be also modified in nuclear matter as $m_K^*=m_K  
\sqrt{\alpha_{\chi SB} / \alpha_2^s} = m_Kf_2\alpha_2^s/f_1$.
Hence we consider the following parametrization   
\begin{equation}
m_K\rightarrow  \frac{f_1}{f_2(\alpha_2^s)^{3/2}}\,(1-C\lambda)m_K,
\label{KmassInNM}
\end{equation}
which has a simple meaning and can be interpreted as 
follows: $C=0$ corresponds to the situation that the kaon properties 
do not change in nuclear matter at all, i.e. $F_K^*m_K^*=F_Km_K$. On
the other hand, if $C\ne 0$, then those of the kaon are linearly varied in
nuclear matter, which is in line with what was observed in
Refs.~\cite{Kaplan:1987sc, Kolomeitsev:1995xz}. Thus, the  
mass term in the effective chiral Lagrangian is changed as 
\begin{align}
F_\pi^2\mathcal{M}\rightarrow F_\pi^{*2}m_\pi^{*2}\left(
\begin{array}{ccc}
1 &0&0\\
0&1 &0\\
0&0&\displaystyle\frac{2F_K^{*2}m_K^{*2}}{F_\pi^{*2}m_\pi^{*2}}-1 
\end{array}
\right),
\end{align}
where 
\begin{align}
F_K^{*}m_K^{*}= F_\pi m_K(1-C\lambda).
\label{comKprop}
\end{align}
In the present work we consider $C$ as an arbitrary external parameter.
In a more consistent approach its value can be adjusted according to the data
on kaon-nucleus scattering or can be related to those on the kaonic atom.
The medium modification in Eq.~(\ref{comKprop})
can be explained in terms of the alteration of the kaon decay
constant and/or of the kaon mass. 

\section{Results and discussions}
\label{sec:ResDis}

\subsection{Density dependence of the low-energy constants} 
\label{subsec:inm-messec}
In order to discuss the density dependence of input parameters in the
mesonic sector according to the definitions in Eq.~\eqref{medpar}, one
should fix the forms of the density-dependent functions. 
We see from Eqs.~\eqref{f1}-\eqref{f3} that at least one of the
density-dependent functions must be adjusted to fit the explicit forms
of the four functions $\alpha_2^s$, $\alpha_4^s$, $\alpha_{\chi SB}$
and $\alpha_2^t$. There are many possible ways of modifying the
functions, since we have only three independent relations between the
four density-dependent functions.  Once the forms of the functions are
fixed, then we can discuss the density dependence of the input
parameters of the model.  Following Ref.~\cite{Yakhshiev:2014hza}, we
try the following form  
\begin{equation}
\alpha_2^s=\exp(-0.65\lambda). 
\label{alpha2s}
\end{equation} 
Then the other three density-dependent functions can be also 
fixed. Moreover, one can fix the form of $\alpha_4^t$ from
Eq.~\eqref{relat}. The parametrization of Eq.~\eqref{alpha2s} is
consistent with the data on low-energy pion-nucleus scattering and
pionic atoms at low densities~\cite{Ericsonbook}. 

\begin{figure}[th]
  \includegraphics[scale=0.23]{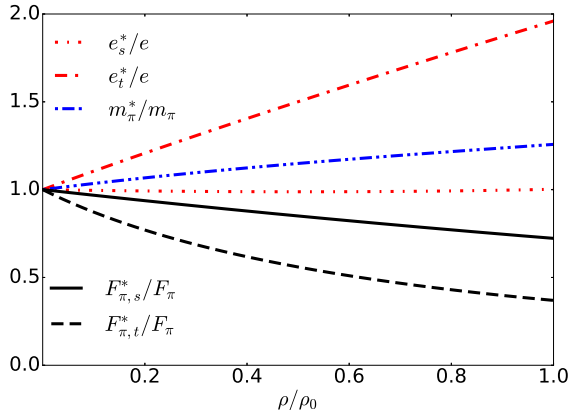}
\caption{ (Color online.) The density dependence of the input parameters
that are defined in Eq.\,(\ref{medpar}). The results shown are those
normalized relatively to their values in free space.} 
\label{Fig1}
  \end{figure}

Now we discuss the density dependence of low-energy constants that 
come into play as the input parameters in the present model. The
results are drawn in Fig.~\ref{Fig1}. 
One can see that the parametrization given in Eq.\,(\ref{alpha2s})
makes the pion mass increased as the density increases. Both the temporal
and spatial parts of the in-medium pion decay constant,  $F_{\pi,t}^*$ and
$F_{\pi,s}^*$, fall off as the density increases (see the solid and
dashed curves). 
These results are in qualitative agreement with those
from chiral perturbation
theory~\cite{Meissner:2001gz,Kirchbach:1997rk} and QCD sum  
rules~\cite{Kim:2003tp}. We refer to Ref.~\cite{Yakhshiev:2013eya} for
a detailed discussion about the consistency of the results in the
present approach and the comparison with other works. 
Nevertheless, we want to note that in contrast to the mentioned
works, the temporal part of the pion decay constant falls off faster
than the spatial one in the present work. 
This comes from the fact that $\alpha_2^s$ is chosen as in
Eq.~\eqref{alpha2s} which is consistent with the data on pionic atoms
only at low  densities~\cite{Ericsonbook}. If one changes the density
dependence of $\alpha_2^s$, then the dependence of the pion decay
constants also  will be altered.

Since the Skyrme parameter $e$ is related to the
$g_{\rho\pi\pi}$  coupling constant, its change in nuclear matter is
deeply related to those of the $\rho$-meson width and its
mass~\cite{Jung:2012sy, Jung:2014jja, Jung:2015piw}.   
Interestingly, $e_s^*$ is almost constant up to normal nuclear matter
density $\rho_0$. This result is a plausible one, because the
in-medium change of $e_s^*$ characterizes how the 
inner core of the skyrmion undergoes the change. On the other hand, 
the spatial part of the pion decay constant $F_{\pi,s}^*$ governs the 
outer shell of skyrmion. Figure~\ref{Fig1} shows that the parameter
$e_s^*$ remains almost constant by the surrounding nuclear environment
up to the normal nuclear matter density (see the dotted curve in
Fig.~\ref{Fig1}).  Further, 
$e_s^*$ starts to increase faster at higher densities. 
At $\rho_0$ we get the result $(F_{\pi,s}^*/F_\pi)^2\approx0.52$, which
crudely explains that the mass contribution from the outer shell of the
soliton in a static approximation is decreased by about
50~\%\footnote{There will be also an inexplicit change due to the
  in-medium modified profile function, which is found to be small.}.
For comparison, we mention that the corresponding contribution from
the inner core remains almost the same, i.e. $(e^*_s/e)^{-2}\approx
1$.\footnote{Note that the contribution from the Skyrme term is
  proportional to the inverse square of the Skyrme parameter.}

The parameter $e_t^*$, which is related to the quantum fluctuations  
of the core of the spinning skyrmion, rises faster, as $\rho/\rho_0$
increases. However, its change in nuclear matter is smaller in
comparison with the density dependence of $F_{\pi,t}^*$. The latter
one is related to the quantum fluctuations of the outer shell of the
skyrmion. The temporal part of the pion decay constant is changed to
be $(F_{\pi,t}^*/F_\pi)^2\approx 0.14$ at $\rho_0$ while the
corresponding temporal part is altered to be as $(e^*_t/e)^{-2}\approx
0.26$.  Thus, in general, we conclude that the outer shell of the
skyrmion is modified larger than its inner core. Figure~\ref{Fig1}
reveals clearly that the temporal parts of the constants change more
strongly in nuclear medium than the spatial ones. These results 
demonstrate that the quantum fluctuations $\sim1/\Omega^{*}$ become
more pronounced in nuclear medium than in free space.  

Concerning the change of the kaon properties, we will regard $C$ as a
free parameter and will examine how it affects the masses of SU(3)
baryons in nuclear matter. 
\subsection{Density dependence of the masses of the 
lowest-lying SU(3) baryons}  
\label{subsec:inm-barsec}

\begin{figure*}[th]
 \includegraphics[scale=0.23]{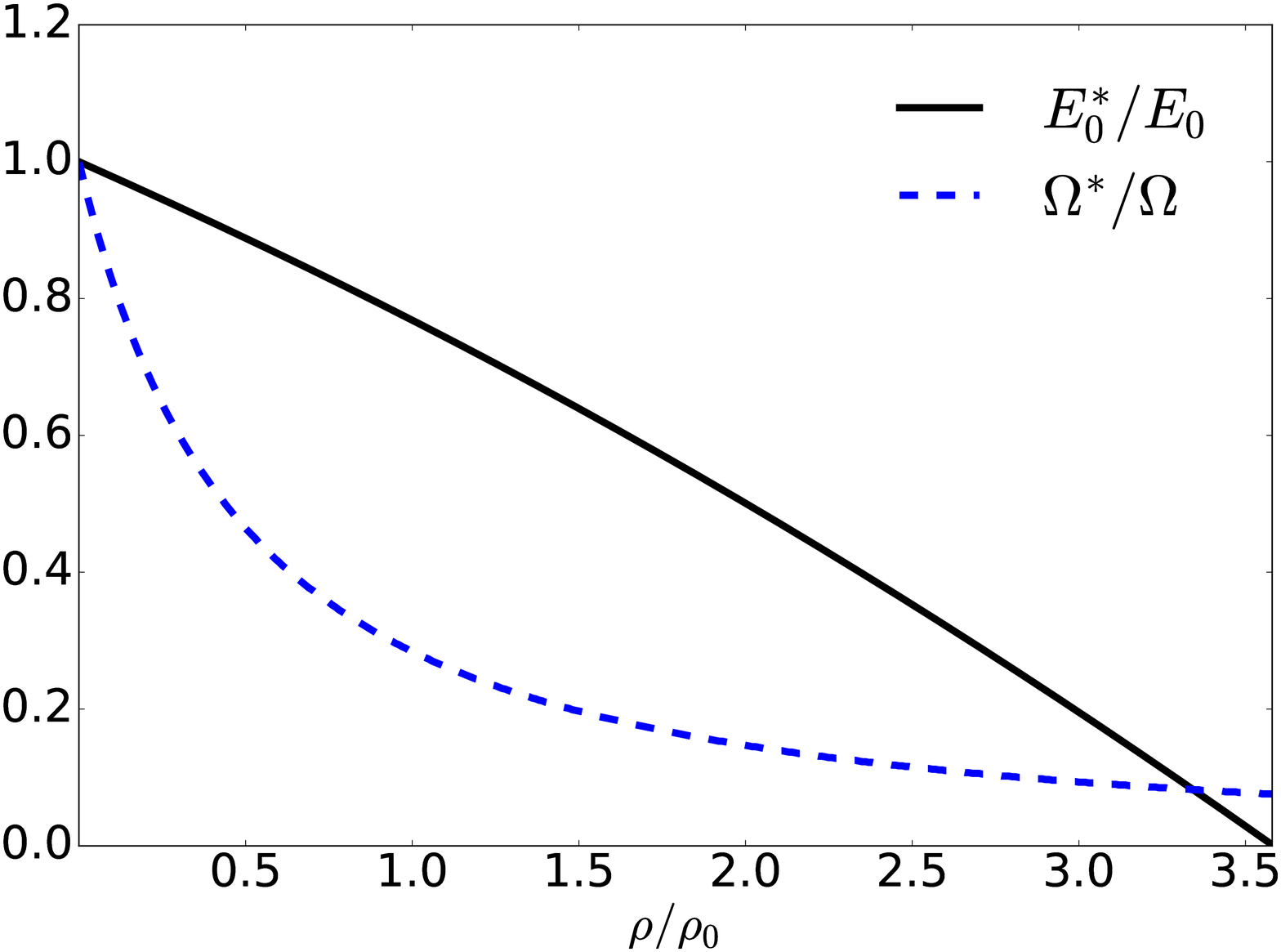}
 \includegraphics[scale=0.23]{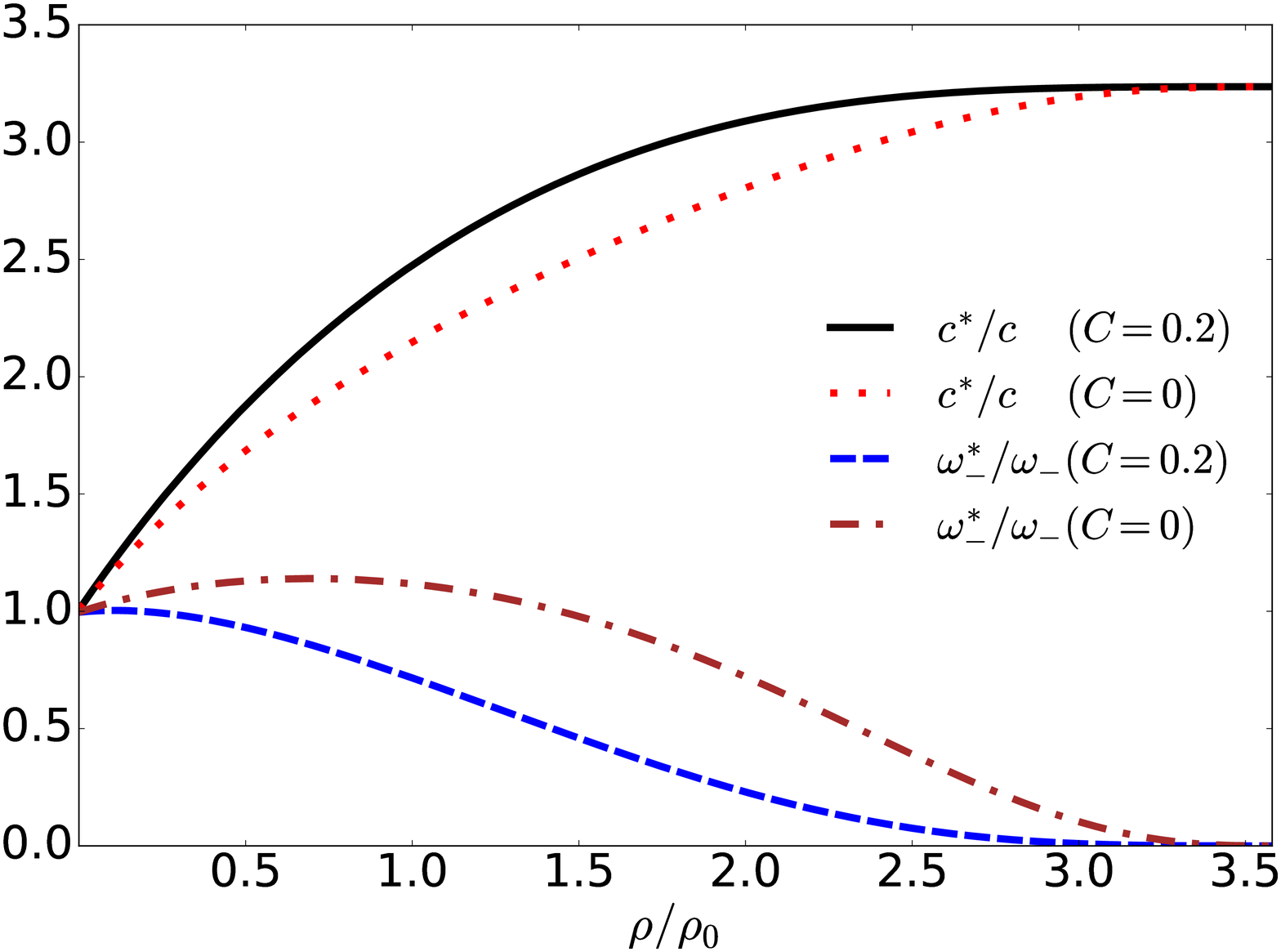}
\caption{ (Color online.) Density dependence of the relative
  density-dependent functions. In the the left panel  $E_0^*/E_0$
  (solid curve) and $\Omega^*/\Omega$ (dashed curve) are drawn as
  functions of $\rho/\rho_0$. They are independent of the kaon  
  properties. In the right panel, $c^*/c$ and $\omega_-^*/\omega_-$
  are depicted as functions of $\rho/\rho_0$ with two different values
  of $C$ ($C=0$ and $C=0.2$) taken into account. The solid and dashed
  curves correspond to the results of $c^*/c$ and
  $\omega_-^*/\omega_-$ with $C=0.2$, respectively, whereas the dotted
  and dot-dashed ones illustrate those of  $c^*/c$ and
  $\omega_-^*/\omega_-$ with $C=0$, respectively. 
} 
\label{Fig2}
\end{figure*}
Since the classical energy of the skyrmion and its moment of
inertia constitute essential parts of the baryon masses, we first
discuss the density dependence of these two quantities. 
The left panel of Fig.~\ref{Fig2} depict the relative classical energy
$E_0^*/E_0$ and the relative moment of inertia $\Omega^*/\Omega$ as
functions of $\rho/\rho_0$. While $E_0^*/E_0$ decreases slowly as the
density increases, $\Omega^*/\Omega$ falls off drastically till $\rho$
reaches a half value of normal nuclear matter density. With the further 
increasing density, $E_0^*/E_0$ decreases in the same manner
and $\Omega^*/\Omega$ start to diminish slowly.  At normal nuclear
matter density, $E_0^*$ is decreased by about $20~\%$. On the other
hand, $\Omega^*$ drops off by about $80\,\%$, which shows that the 
rotational $1/N_c$ corrections increases as the density
increases. As a result, the nuclear matter becomes stabilized  
around the saturation density
$\rho_0$. At higher densities, these functions describe the stiffness of
the equations of state for nuclear matter. 
Note that these two quantities, $E_0$ and $\Omega$, are not at all
influenced by the change of the kaon properties.  
The consequence of this behavior will be discussed soon. 

Concerning the parameters in the strangeness sector, i.e. $c^*$ and
$\omega_-^*$, we will present the results for two different cases: 
\begin{itemize}
\item[{i)}] We do not change the kaon properties in nuclear
  matter, i.e. $F_\pi^* m_K^*=F_\pi m_K$ or  $C=0$.
\item[{ii)}] We make $F_\pi^* m_K^*$ decreased linearly as
  the density of nuclear matter increases. This corresponds to the
  value $C=0.2$.  
\end{itemize}
By doing this, we can see how the change of the kaon properties affect
the mass shift of the SU(3) baryons. In the right panel of Fig.~\ref{Fig2},
the results of $c^*/c$ and $\omega_-^*/\omega_-$ are depicted as
functions of $\rho/\rho_0$ with the above-mentioned two different cases 
considered. When we turn on the value of $C$, $c^*/c$ increases faster
than that with $C=0$. The behavior of $\omega_-^*/\omega_-$ is also
changed when $C=0.2$ is taken. If one switches off $C$,
$\omega_-^*/\omega_-$ starts to increase first and then falls off
slowly, as the density increases. However, when one uses $C=0.2$,
$\omega_-^*/\omega_-$ drops off monotonically, which is distinguished
from the case with $C=0$. As will be shown below, this change with the
finite value of $C$, i.e. the change of the kaon properties, will have
a clear effect on the mass splitting of the baryon octet in nuclear
matter. The physical meaning of $\omega_-^*$ is the quantum
fluctuation along the strangeness direction. So, it plays an essential
role in determining the hyperon masses as shown in
Eq.~\eqref{massformula}.  On the other hand, $c^*$ is related to the
isospin splitting within the same multiplet when the strangeness is
equal to zero. Of course it provides a certain contribution to the
hyperon masses (see Eq.~\eqref{massformula}).  

\begin{table}[hbt]
\caption{Values of the density-dependent skyrmion functionals at normal nuclear
  matter density $\rho_0$ in comparison with those in free space.
Those of the functionals with nonzero strangeness are presented
with the two different values of $C$ taken into account. }
 \begin{ruledtabular}
\begin{tabular}{c|c|cc}
 Skyrmion & Free space  &\multicolumn{2}{c}{Values at $\rho=\rho_0$}\\
 \cline{3-4}
 functionals & values&$C=0$&$C=0.2$\T \\
 \hline
  $E_0^{*}$[MeV] &$865.60$ &$665.04$ & $665.04$\T \\
 $\Omega^{*}$[MeV$^{-1}$] & $5.116\times10^{-3}$  &
  $1.453\times10^{-3}$ & $1.453\times10^{-3}$\\
 $\Phi^{*}$ [MeV$^{-1}$] &$1.852\times10^{-3}$  &
  $5.000\times10^{-4}$  & $5.000\times 10^{-4}$\\
 $\Gamma^{*}$ [MeV$^{-1}$] & $3.995\times10^{-3}$  &
  $5.442\times10^{-3}$  & $5.442\times10^{-3}$\\
 $\omega_{-}^{*}$ [MeV] &202.44  & 226.03  & 144.53 \\
 $c^{*}$ & 0.309  & 0.664 & 0.765\\
\end{tabular}
\end{ruledtabular}
\label{table1}
\end{table}
In Table~\ref{table1}, we list the values of the density-dependent
skyrmion functionals at normal nuclear matter density $\rho_0$, comparing them
with those in free space. As mentioned already, $E_0^*$ and $\Omega^*$
are reduced approximately by $20\,\%$ and $80\,\%$, respectively, at
$\rho_0$ in comparison with the corresponding values in free
space. Note that the functionals $\Phi^*$ and $\Gamma^*$ have no
explicit influence on the baryon masses but they influence the values
of the other two functionals $\omega_-^*$ and $c^*$. Moreover, they do
not depend on $C$. When $C$ is turned off, the value of $\omega_-^*$
increases by about $12\,\%$ at $\rho_0$, compared with that in free
space. However, if one considers the in-medium changes of the kaon
properties by taking $C=0.2$, $\omega_-^*$ is reduced by about
$29\,\%$. On the other hand, $c^*$ increases for both cases.  
As expected, the changes of the kaon properties in nuclear matter
indeed influence the quantities in the strangeness direction and will
consequently affect the characteristics of the hyperons in nuclear
matter.  

In Table~\ref{table2}, we list the results of the masses of the
baryon octet and decuplet both in free space and in nuclear
matter at $\rho_0$.  The values of the nucleon mass in free space and
in nuclear matter at $\rho_0$, and the mass of the $\Delta$-isobar in
free space are used as input parameters of the model. 
The masses of the nucleon and $\Delta$ in free space are employed to
fix the values of pion decay constant and skyrme parameter in free
space. The in-medium mass of nucleon at $\rho_0$ fixes the
one of the density-dependent functions $f_{1,2,3}$.

\begin{table}[hbt]
\caption{Results of the masses of the baryon octet and decuplet both
  in free space and in nuclear matter at $\rho_0$ in units of
  MeV. Note that the masses of the nucleon and the $\Delta$ isobar are
  used as input, which are marked by asterisk (`*') as the
  superscripts of the corresponding numbers.} 
\begin{ruledtabular}
\begin{tabular}{c|cccc}
Baryon& Experimental& Free space &\multicolumn{2}{c}{Mass at $\rho=\rho_0$}\\
 \cline{4-5}
&mass&mass &$C=0$&$C=0.2$\T \\
 \hline
 $N$ & 939  & 939$^*$  & 923$^*$ & 923$^*$\T\\
 $\Lambda$& 1115  &1075  & 1004 & 960\\
 $\Sigma$&1189  &1210  & 1236 & 1122\\
 $\Xi$ & 1315  & 1302  & 1221 & 1088\\
\hline
 $\Delta$ & 1232  & 1232$^*$  & 1956 & 1956\T\\
 $\Sigma^{*}$  &1385  & 1301  & 1921 & 1912\\
 $\Xi^{*}$& 1530  & 1392  & 1906& 1878\\
 $\Omega$  &1672 &1508  &1911 & 1854\\
 \end{tabular}
\end{ruledtabular}
\label{table2}
\end{table}

We find that in general the masses of the baryon octet tend to
decrease in nuclear matter except for that of $\Sigma$ which increases
with $C=0$ but drops off with $C=0.2$ considered. The mass of the
$\Lambda$ is changed as $m_\Lambda^*/m_\Lambda\approx 0.93$ for $C=0$, 
whereas $m_\Lambda^*/m_\Lambda\approx 0.89$ for $C=0.2$. 
It is interesting to compare the present results with that 
from SU(3) chiral effective field theory~\cite{Petschauer:2015nea} in
which $m_\Lambda^*/m_\Lambda\approx 0.73$ was obtained. The mass of
the $\Xi$ hyperon is changed in a similar manner:
$m_\Xi^*/m_\Xi\approx 0.94$ for $C=0$ and $m_\Xi^*/m_\Xi\approx 0.84$
for $C=0.2$, respectively, both of which are more reduced in nuclear
matter in comparison with that from the quark-meson coupling model
with the bag radius of the free nucleon $R_0=0.8$\,fm,
i.e. $m_\Xi^*/m_\Xi\approx  0.98$~\cite{Saito:1994kg}. Thus, the
present results of the $\Lambda$ and $\Xi$ mass dropping are in
qualitatively agreement with those from the other approaches. 

In contrast with the masses of the baryon octet, those of the decuplet
are increased drastically, as the density of nuclear matter
increases. This can be understood from Eq.~\eqref{massformula}. The
second term of Eq.~\eqref{massformula} makes the baryon decuplet split from
the octet. As shown already in the left panel of Fig.~\ref{Fig2}, the
moment of inertia $\Omega^*$ drops off rapidly as the density of
nuclear matter increases, which makes the second term of 
Eq.~\eqref{massformula} increase very fast. This brings about the
drastic increment of the spin-3/2 hyperon masses. When $C=0.2$ is
used, the masses of the hyperon decuplet still increase but are found
to be smaller than the case with $C=0$.  
 
Theoretically, it is of more interest to study the density effects
on the mass splittings of the hyperons, since soliton models predict
them quantitatively in comparison with the experimental data. 
We first express the formulae for the mass splittings of the hyperon
octet, given as 
  \begin{align}
    m_{\Sigma}^*-m_{\Lambda}^*&=\frac{1-c^*}{\Omega^*},
    \label{mSigma-mLambda}\\
    m_{\Xi}^*-m_{\Sigma}^*&=\omega^*_{-}+\frac{5(c^*-1)(c^*+1)}{8\Omega^*},\\
    m_{\Lambda}^*-m_{N}^*&=\omega^*_{-}+\frac{3(c^*-1)(c^*+1)}{8\Omega^*},
  \end{align}
and the hyperon decuplet, written by
  \begin{align}
    m_{\Sigma^{*}}^*-m_{\Delta}^*&=
    \omega^*_{-}+\frac{(c^*-1)(3c^*+7)}{8\Omega^*},\\
    m_{\Xi^{*}}^*-m_{\Sigma^{*}}^*&=\omega^*_{-}+\frac{5(c^*-1)(c^*+1)}{8\Omega^*},\\
    m_{\Omega}^*-m_{\Xi^{*}}^*&=\omega^*_{-} + \frac{(c^*-1)(7c^*+3)}{8\Omega^*}
    \label{mOmega-mXi}
  \end{align}
in nuclear matter. 

  \begin{figure*}[th]
  \includegraphics[scale=0.23]{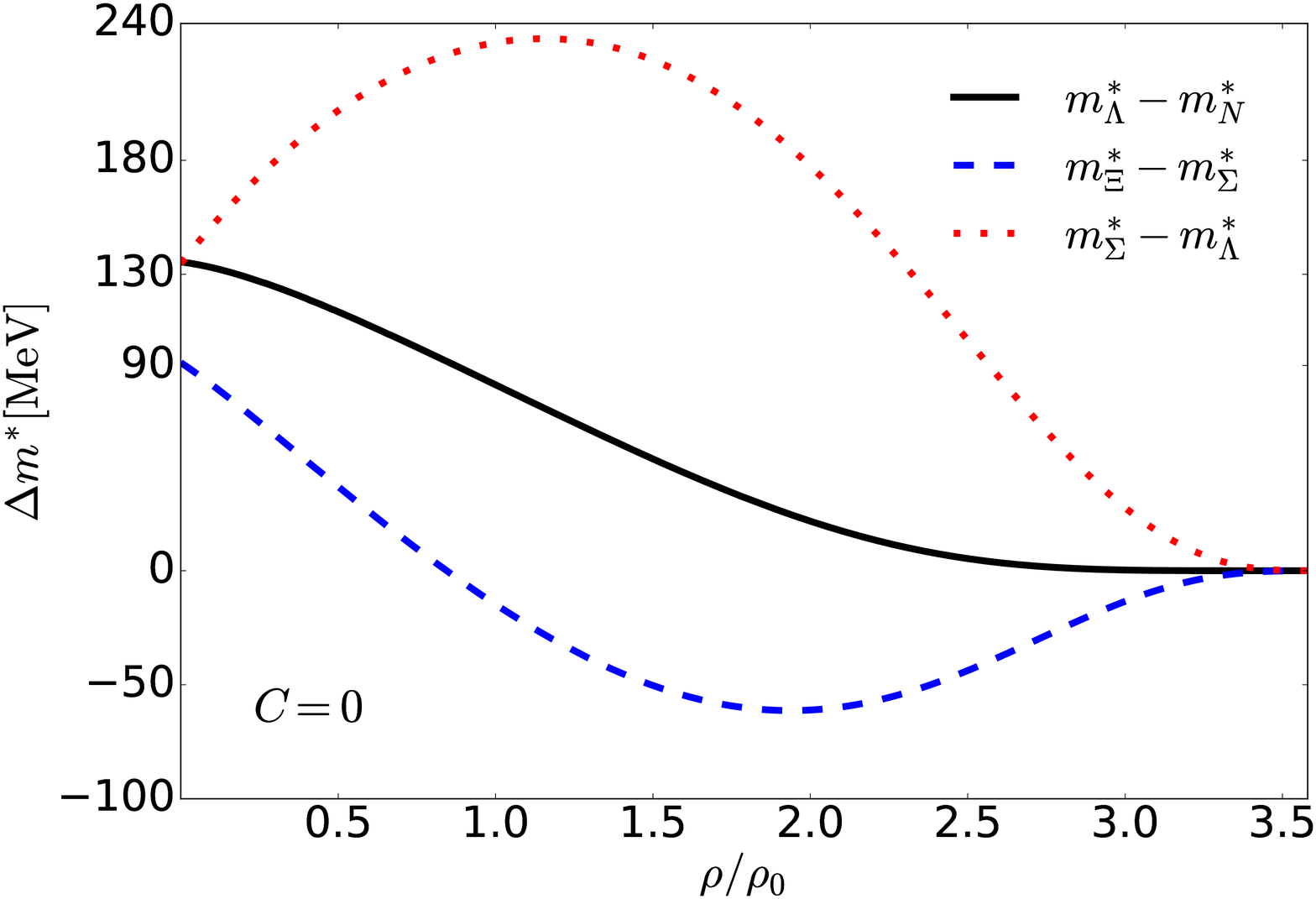}
  \includegraphics[scale=0.23]{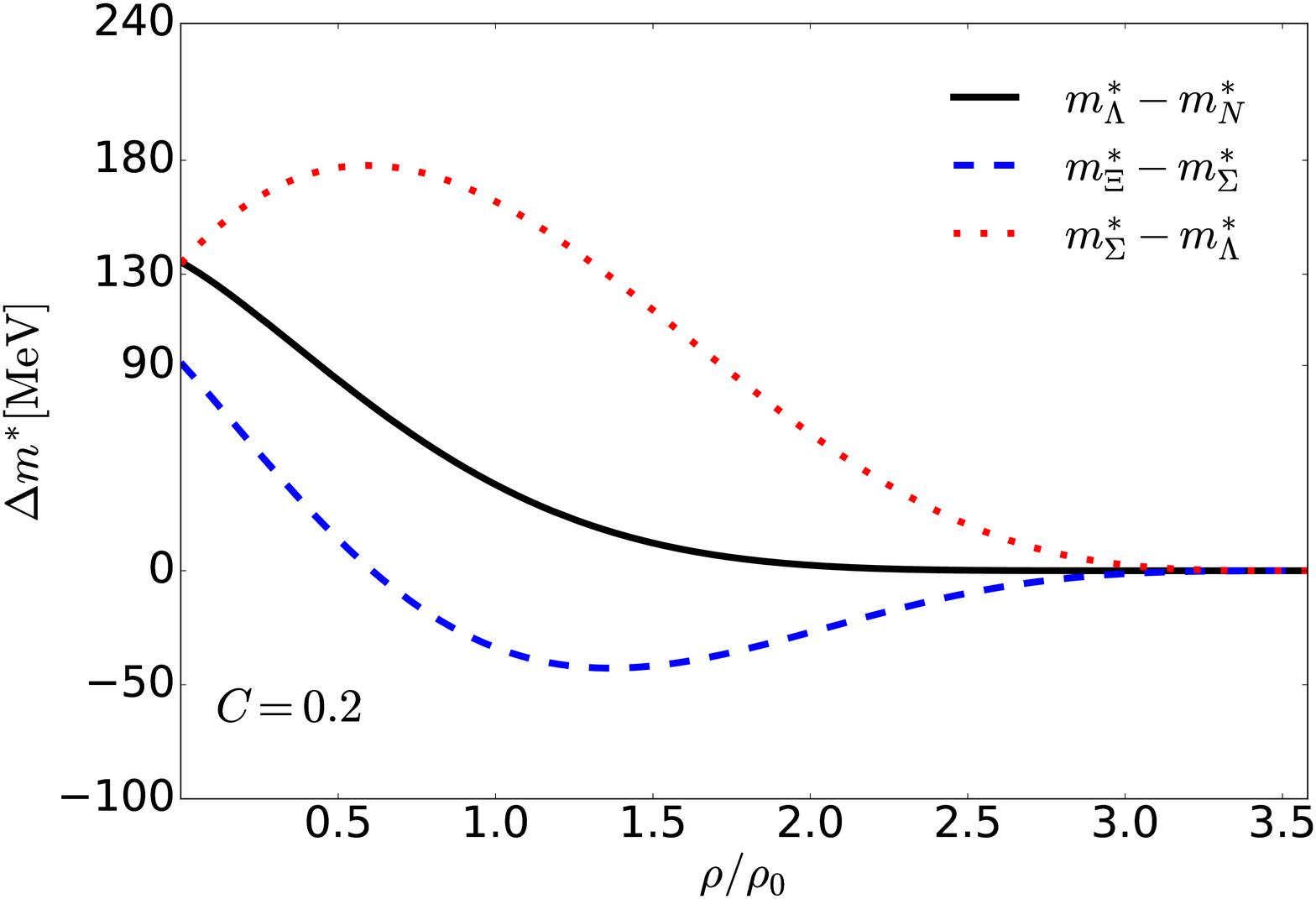}
\caption{(Color online.) The density dependence of the mass splittings
  of the baryon octet. In the left panel, the results of
  $m_\Lambda^*-m_N^*$, 
  $m_\Sigma^*-m_\Lambda^*$, and $m_\Xi^*-m_\Sigma^*$ are
  illustrated by the solid curve, the dotted one, and the dashed one,
  respectively, with $C=0$ considered. In the right panel, those with
  $C=0.2$ are drawn. The notations are the same as in the left panel.}  
\label{Fig3}
  \end{figure*}
In the left panel of Fig.~\ref{Fig3} the mass splittings of the
hyperon octet are drawn without changing the kaon properties in
nuclear matter, whereas in its right panel those are depicted with
$C=0.2$ used. The general density dependence of the mass splittings is
not much changed by introducing the finite $C$, one can clearly see
that the magnitude of the mass splittings is reduced when $C=0.2$ is
employed. As seen in Fig.~\ref{Fig3}, the result of
$m_{\Sigma}^*-m_{\Lambda}^*$ illustrated in the dotted curve exhibits a
different density dependence. Since the $\Sigma$ and $\Lambda$ have
the same strangeness, $\omega_-^*$ is not involved in this
splitting. As shown in Eq.~\eqref{mSigma-mLambda},
$m_{\Sigma}^*-m_{\Lambda}^*$ is proportional to $1-c^*$ and $1/\Omega^*$. As
the density increases, both the numerator and denominator start to
decrease but $\Omega^*$ falls off much faster. Thus, the mass
splitting $m_{\Sigma}^*-m_{\Lambda}^*$ grows larger until the density
reaches $\rho\approx 1.2\rho_0$ ($\rho\approx 0.7\rho_0$) in the case
of $C=0$ ($C=0.2$),  and then it drops off till $\rho\approx
3.5\rho_0$ is reached.  

The mass splitting $m_\Lambda^* - m_N^*$ falls off monotonically as the
density increases, whereas $m_\Xi^* - m_\Sigma^*$ lessens till
$\rho\approx 2\rho_0$ with $C=0$ ($\rho\approx 1.5\rho_0$ with
$C=0.2$). Note that all the masses of baryon octet become
degenerate when the density reaches $\rho \approx 3.5\rho_0$. It
implies that the SU(3) flavor symmetry is restored around $3.5\rho_0$
within the present framework. Interestingly, if we include the
in-medium changes of the kaon properties, it comes about the
degeneracy of the masses at lower densities. 

   \begin{figure*}[htb]
 \includegraphics[scale=0.23]{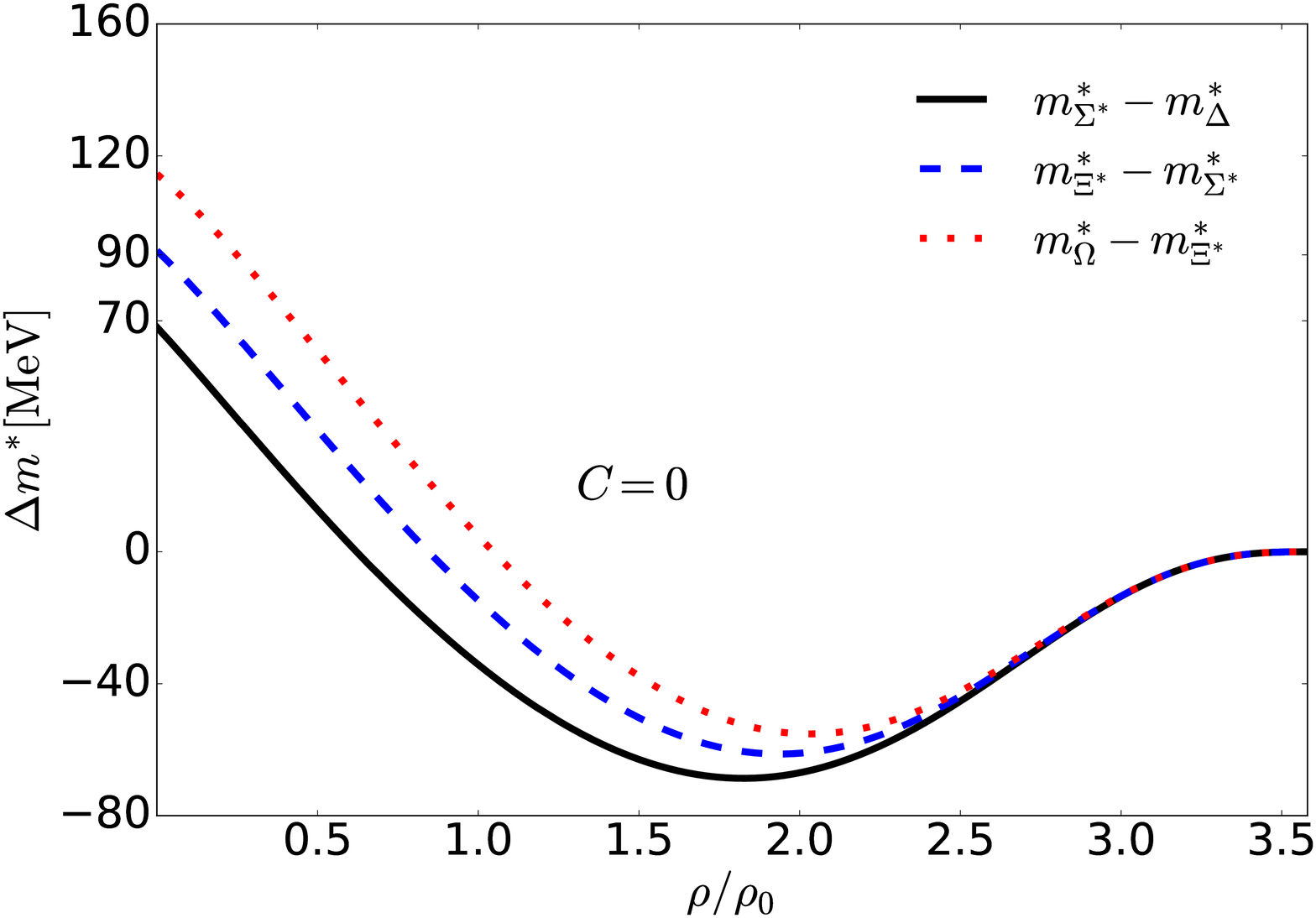}
  \includegraphics[scale=0.23]{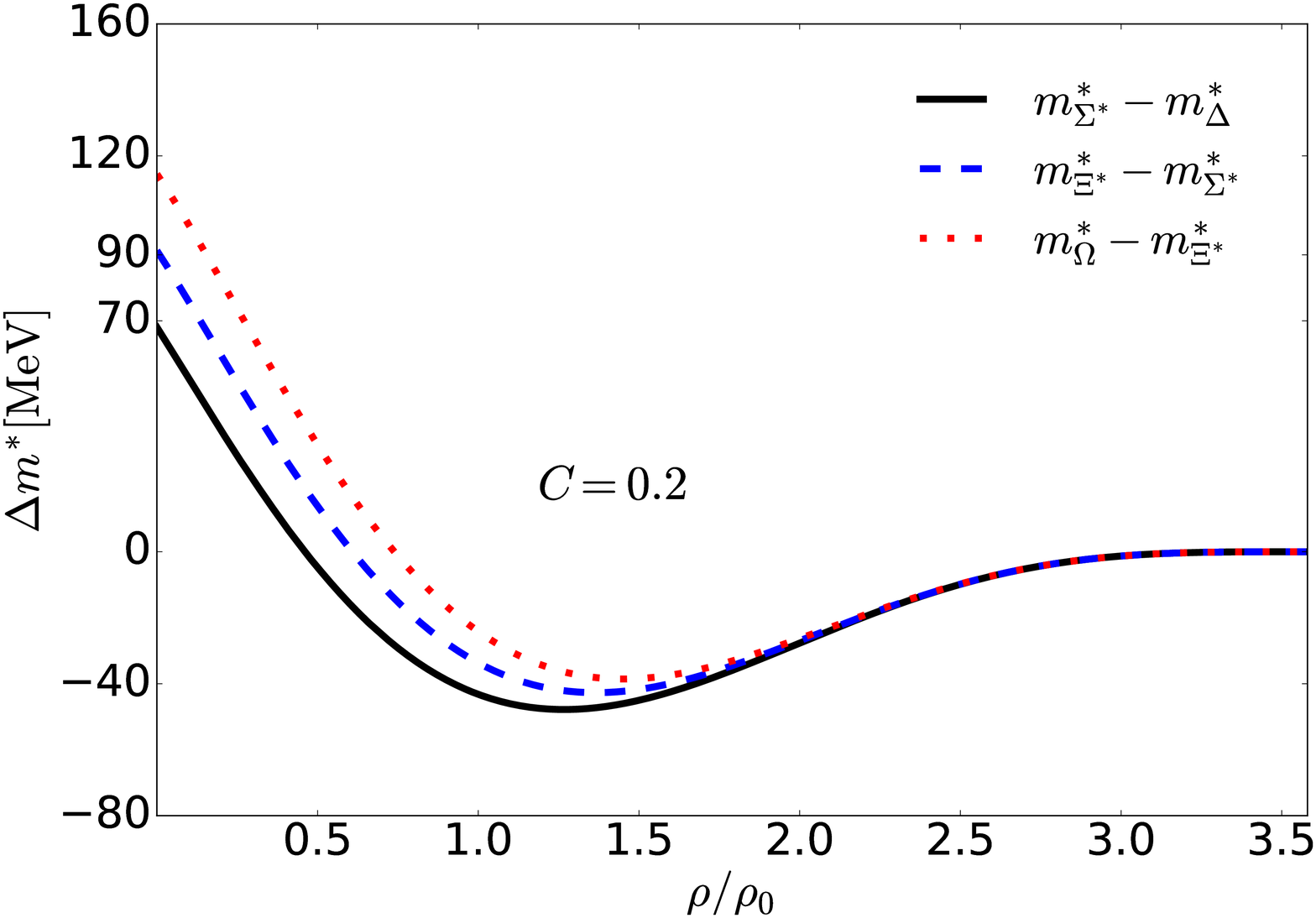}
\caption{(Color online.) The density dependence of the mass splittings
  of the baryon decuplet. In the left panel, the results of
  $m_{\Sigma^*}^*-m_\Delta^*$, $m_{\Xi^*}^*-m_{\Sigma^*}^*$, and
  $m_\Omega^*-m_{\Xi^*}^*$ are illustrated by the solid curve, the dotted
  one, and the dashed one, respectively, with $C=0$ considered. In the
  right panel, those with $C=0.2$ are drawn. The notations are the
  same as in the left panel.} 
\label{Fig4}
  \end{figure*}  
Figure~\ref{Fig4} represents the numerical results of the mass
splittings of the baryon decuplet. The general tendency of the results
is in line with that of the $m_\Xi-m_\Sigma$ shown in
Fig.~\ref{Fig3}. This can be understood by examining the formulas
given in Eqs.~\eqref{mSigma-mLambda} and \eqref{mOmega-mXi}.
Interestingly, there is an identity 
\begin{align}
m_\Xi-m_\Sigma  = m_{\Xi^*}-m_{\Sigma^*}  
\end{align}
which is kept in nuclear matter too.
All other mass splittings of the baryon decuplet exhibit similar
behaviors as the density increases. Compared the case of the baryon
octet, the degeneracy takes place at lower densities.
  
\section{Summary and outlook}
\label{sec:Sum&Out}

In this work we investigated the density dependence of the baryon octet
and decuplet masses in nuclear matter within the framework of the 
in-medium modified SU(3) Skyrme model. For simplicity we first
concentrated on the medium modifications arising from the in-medium
changes of the pion properties, which encodes the modification of the pion
propagation in nuclear matter. The parameters were determined by
describing the properties of nuclear matter near the saturation point
$\rho_0$. In particular, the in-medium modified meson parameters 
provide the equation of states in the wide range of nuclear matter
densities. In addition to this, we introduced the changes of the
\emph{produced} kaon properties in nuclear matter, which are in line
with Refs.~\cite{Kaplan:1987sc, Kolomeitsev:1995xz} and examined their 
effects on the masses of the baryon octet and decuplet. 

We discussed also that the changes of the mesonic properties are 
generally in qualitative agreement with those from in-medium chiral
perturbation theory~\cite{Meissner:2001gz,Kirchbach:1997rk}  and the
QCD sum rules~\cite{Kim:2003tp} except for the relative density
dependencies of $F_{\pi,s}^*$ and $F_{\pi,t}^*$.  The present results
of the SU(3) baryon masses in nuclear matter are also in qualitative
agreement with those from in-medium chiral effective field
theory~\cite{Petschauer:2015nea} and quark-meson coupling
model~\cite{Saito:1994kg}.  

In order to study the effects of the modified kaon properties, we have
to go beyond the present simple scheme. We need to associate with kaon
dynamics in nuclear matter in close relation with experimental data on
kaon-nucleus scattering and kaonic atoms. It is also of great
importance to investigate the equation of states for strange matter with
regards to the interior structure of neutron stars. The relevant
studies are under way. 

\newpage

\begin{acknowledgments}
The authors are grateful to P.~Gubler, A.~Hosaka, T.~Maruyama and
M.~Oka for useful discussions.  The authors want to express their
gratitude to the members of the Advanced Science Research Center at
Japan Atomic Energy Agency for the hospitality during their visit, 
where part of the
present work was done. The work is supported by Basic Science Research
Program through the National Research Foundation (NRF) of Korea funded
by the  Korean government (Ministry of Education, Science and
Technology, MEST), Grant No. 2016R1D1A1B03935053 (UY) and Grant  
No. NRF-2018R1A2B2001752 (HChK). 
\end{acknowledgments}

\end{document}